\documentstyle{article}
\newcommand{\bm}[1]{\mbox{\boldmath$#1\!$}}
\newcommand{\bn}[1]{\mbox{\boldmath$#1$}}

\newcommand{\beq}{\begin{equation}}
\newcommand{\eeq}{\end{equation}}
\newcommand{\bea}{\begin{eqnarray}}
\newcommand{\eea}{\end{eqnarray}}

\begin{document}
\large{
\baselineskip=25pt

{\Huge{\centerline{\bf The influence of a metallic sheet on }}}
{\Huge{\centerline{\bf an evanescent mode atomic mirror}}}
\vskip 1.5cm
\centerline{{\bf J. B. Kirk$^{*}$, C. R. Bennett$^{\dagger}$, M. Babiker$^{\dagger}$} and {\bf S. Al-Awfi}$^{\dagger\dagger}$}
\centerline{$^{*}$ Department of Physics, University of Essex, Colchester, Essex CO4 3SQ, UK}
\centerline{$^{\dagger}$ Department of Physics, University of York, Heslington, York, YO10 5DD, UK}
\centerline{$^{\dagger\dagger}$ Department of Physics, King Abdulaziz University, P O Box 344, Medina, Saudi-Arabia}
\vskip 2.0cm
\section*{Abstract}

A theory of evanescent mode atomic mirrors utilising a metallic sheet on a dielectric
substrate is described. The emphasis here is on the role of the metallic sheet 
and on the evaluation of atomic trajectories using the field-dipole orientation picture.  At low intensity, the atomic reflection process 
is controlled by two separate mechanisms both of which are modified by the presence of the metallic sheet
and influenced by the use of the field-dipole orientation picture.  The first mechanism involves the spontaneous force which
accelerates the atom parallel to the sheet plane. The second mechanism involves 
the combined dipole plus van der Waals force which acts to repel the atom from the surface, decelerating its motion until it attains 
an instantaneous halt before changing direction away from the surface at an appropriate turning point in the trajectory.  
Various quantitative features arising from varying the controllable parameters of the system, including screening effects as well 
as desirable enhancement effects, are pointed out and discussed.  
\vskip 1.0cm

\centerline{\bf PACS numbers: 42.50.Vk, 32.70.Jz, 32.80.Pj, 32.80.Lg}
\newpage
 One of the primary aims, yet to be fully accomplished in the rapidly developing field of atom optics
 is the standardisation of the atomic mirror as a routine atom optical element. The idea of an atomic mirror,  a device capable of reflecting atoms, 
was first put forward by Cooke and Hill in 1982 [1].  Since then,  a number of studies, both theoretical and experimental, have sought to explore various aspects of atomic mirrors,
based on different mechanisms of
 atom reflection from planar surfaces [2-19].  In particular, the evanescent mode atomic mirror is suitable for neutral atoms undergoing
  electric dipole transitions [14,18,19].
 The mechanism in this type of mirror makes use of light 
 evanescing into the vacuum region outside the planar surface of a dielectric when laser light is internally  
 reflected.  The evanescent light sets 
 up a repulsive dipole potential which acts on any neutral atom possessing a transition frequency 
 at near resonance with, and blue-detuned from, the laser frequency.  The limiting
factors of such a mirror stem primarily from heating effects and also from fluctuations due to the light
which become negligible at low intensities.  A device operating at low intensities
was thus needed.  This was the motivation behind the studies which  highlight the role of a thin metallic layer
deposited on the surface 
 in providing an enhancement of the evanescent component [18,19].

 The purpose of this article is to report a theory of evanescent 
mode atomic mirrors  with a metallic sheet, exploring the nature of the enhancement, determining the general features of the mirror and evaluating the 
trajectories for a typical set of parameters.
 
 It turns out that a quantitative theoretical analysis demands the implementation of a number of steps. 
 First, a mode normalisation  procedure in terms of intensity is needed for the incident light mode responsible for the generation of the evanescent component. 
Secondly, the jump conditions should be implemented involving the surface current arising from the finite two-dimensional conductivity of the metallic sheet.  
Thirdly, the attractive force between the atom and the dielectric in the presence of the metallic
sheet should be included in the dynamics to determine the trajectories.  Finally,  use should made of the field-dipole orientation 
picture for evaluating the radiation forces as well as the van der Waals-type atom-surface force.  The role of the field-dipole orientation picture is to determine an average
value for the local orientation of the electric dipole moment vector in the presence of the
evanescent field.  We show here that a programme 
incorporating the above theoretical features permits the atomic trajectories to be determined by direct solution of the equation of motion, leading either 
to a reflection of the atom off the mirror or a collision with it, in a manner dependent on the chosen set
 of parameters.  Existing treatments determining trajectories are primarily based on Monte Carlo techniques.
 
 The basic elements comprising the atomic mirror are shown schematically
 in Fig. 1.  Here a metallic sheet in the form of an infinitesimally thin layer is deposited on 
 the planar surface of a dielectric substrate (or glass prism).  Light of frequency $\omega$
 incident from within the dielectric is internally reflected at the interface 
 between the dielectric substrate and the metallic sheet.  This creates a field in the vacuum
 region which is decaying with distance from the metallic sheet and is propagating 
 along the surface.  A neutral atom possessing a transition frequency $\omega_{0}<\omega$ and
 which is moving in the plane of incidence would be subject to the repulsive dipole force plus an attractive atom-surface force
 and will also experience a light pressure force parallel to the surface.  
 As we show below the combined influence of these forces can be
 made to control the reflection process.
 
 The relevant electric field vector at frequency $\omega$ can be written 
in terms of incident (I), reflected (R) and  evanescent (1) parts as follows
 \beq
 {\bf E}({\bf },{\bf r},t)=\left\{\left({\bf E}_{I}({\bf k}_{\parallel},{\bf r},t)
 +{\bf E}_{R}({\bf k}_{\parallel},{\bf r},t)\right)\theta(-z)+{\bf E}_{1}({\bf k}_{\parallel},{\bf r},t)\theta(z)\right\}
 a+h.c. \label{0}
 \eeq
 where $\theta$ is the unit step function and the fields are given by
 \beq
 {\bf E}_{I}({\bf k}_{\parallel},{\bf r},t)=A_{I}\left(1,0,-\frac{k_{\parallel}}{k_{z2}}\right)
 e^{ik_{z2}z}e^{i({\bf k_{\parallel}{\bf .r}_{\parallel}}-\omega t)}\label{1}
 \eeq
 \beq
 {\bf E}_{R}({\bf k}_{\parallel},{\bf r},t)=A_{R}\left(1,0,\frac{k_{\parallel}}{k_{z2}}\right)
 e^{-ik_{z2}z}e^{i({\bf k_{\parallel}{\bf .r}_{\parallel}}-\omega t)}\label{2}
 \eeq
 \beq
 {\bf E}_{1}({\bf k}_{\parallel},{\bf r},t)=B\left(1,0,\frac{ik_{\parallel}}{k_{z1}}\right)
 e^{-k_{z1}z}e^{i({\bf k_{\parallel}{\bf .r}_{\parallel}}-\omega t)}\label{3}
 \eeq
 Here ${\bf k}_{\parallel}$ is the wavevector parallel to the surface.  Its magnitude $k_{\parallel}$ is given by 
 $c^{2}k^{2}_{\parallel}=\omega^{2}\epsilon_{2}\sin^{2}\phi$ where $\phi$ is the angle of incidence. The three quantities between the 
brackets in each of Eqs.(\ref{1}) to (\ref{3}) stand for the vector components parallel to ${\bf k}_{\parallel}$,  perpendicular to it on the surface plane and along the z-direction, respectively. 
 $k_{z1}$ and $k_{z2}$ (both real) are defined by
 \beq
 k_{z1}^{2}=k_{\parallel}^{2}-\frac{\epsilon_{1}\omega^{2}}{c^{2}}>0;\;\;\;\;\;
 k_{z2}^{2}=\frac{\epsilon_{2}\omega^{2}}{c^{2}}-k_{\parallel}^{2}>0
 \eeq
Finally,  $A_{I}, A_{R}$ and $B$ are field amplitude factors, to be determined.  The notation is such 
that parameters associated with the substrate are
 labelled by the subscript 2,  while for the outer region (vacuum) the label is 1.  Both dielectric functions  $\epsilon_{1}$ and 
 $\epsilon_{2}$ are assumed to be frequency-independent and we take $\epsilon_{1}=1$,  as appropriate for vacuum.  
 The position vector is written as ${\bf r}=({\bf r}_{\parallel},z)$
 in terms of an in-plane position vector ${\bf r}_{\parallel}$ and a z-coordinate relative to the metallic sheet. 
 The role of the metallic sheet is primarily to provide a two-dimensional
 charge density $n_{s}$ and, so, an electrical conductivity $in_{s}e^{2}/m^{*}(\omega+i\gamma)$ where $m^*$ and $e$
are the electronic effective mass and charge and $\gamma\ll \omega$ accounts for metallic plasma loss effects.  The metallic sheet only enters the formalism via the 
 electromagnetic boundary conditions involving the tangential components of the magnetic fields corresponding to Eqs.(\ref{1}) 
 to (\ref{3}),  which can be calculated using Maxwell's equation
 ${\bf H}=-(i\epsilon_{0}c^{2}/\omega){\bm {\nabla}}\times{\bf E}$.  Application of the first boundary condition, namely the continuity of the tangential
 component of the electric field vector at $z=0$, yields
 \beq
 A_{I}+A_{R}=B\label{6}
 \eeq
 The second electromagnetic boundary condition is that the tangential component of the magnetic field vector experiences a discontinuity
 at $z =0$ arising from the surface current induced by the in-plane component of the electric field at the metallic sheet.  We have
 \beq
 H_{\parallel}(0_{-})-H_{\parallel}(0_{+})=\frac{in_{s}e^{2}}{m^{*}(\omega+i\gamma)}E_{\parallel}(0)
 \eeq
 where $0_{\pm}$ are the limits as $\xi\rightarrow 0$ of ($0\pm\xi$). Application of this boundary condition leads to a second relation connecting the field amplitudes
 \beq
 \frac{\epsilon_{2}}{k_{z2}}\left(A_{I}-A_{R}\right)+i\frac{\epsilon_{1}}{k_{z1}}B=\frac{in_{s}e^{2}}
{\epsilon_{0}m^{*}\omega(\omega+i\gamma)}\label{8}
 \eeq
 Elimination of $A_{R}$ between Eqs.(\ref{6}) and (\ref{8}) yields straightforwardly
\beq
B=2A_{I}\left[i\frac{k_{z2}}{\epsilon_{2}}\left(\frac{n_{s}e^{2}}{\epsilon_{0}m^{*}\omega(\omega+i\gamma)}
-\frac{\epsilon_{1}}{k_{z1}}\right)+1\right]^{-1}\label{9}
\eeq 
The next step is to determine the value for the amplitude $A_{I}$.  This is the amplitude of the
incident field in the unbounded bulk of material 2 and the value of $A_{I}$ is such that the Hamiltonian ${\cal H}_{I}$ reduces to the 
canonical form
\beq
{\cal H}_{I}=\frac{\epsilon_{0}}{2}\int\;d{\bf r}\left\{\epsilon_{2}E_{I}^{2}+\frac{1}{(\epsilon_{0}c)^{2}}H_{I}^{2}\right\}=\frac{1}{2}\hbar\omega
\left(aa^{\dagger}+a^{\dagger}a\right) 
\eeq
Thus we find
\beq
A_{I}^{2}=\frac{\hbar k_{z2}^{2}c^{2}}{2V\epsilon_{0}\epsilon_{2}^{2}\omega}
\eeq
where $V$ is the (large) volume of material 2.  

The average radiation force acting on an atom 
of transition frequency $\omega_{0}$ moving in the vacuum region at velocity ${\bf v}$ is given by the well known expression [20,22]
\beq
F({\bf r,v})=2\hbar\left\{\frac{\Gamma\Omega_{R}^{2}{\bf k}_{\parallel}-\frac{1}{2}\Delta{\bn {\nabla}}\Omega_{R}^{2}}{\Delta^{2}
+2\Omega_{R}^{2}+\Gamma^{2}}\right\}={\bf F}_{s}+{\bf F}_{d}\label{12}
\eeq
where ${\bf F}_{s}$ corresponds to the first term,  identified as the spontaneous force along the wave propagation direction 
${\bf k}_{\parallel}$, and ${\bf F}_{d}$ corresponds to the second term,  identified as the dipole force; 
$\Delta$ is the dynamic detuning given by
\beq
\Delta=\Delta_{0}-{\bf k}_{\parallel}{\bf .v}
\eeq
with $\Delta_{0}=\omega-\omega_{0}$  the static detuning of the light from the atomic resonance.  The force given in Eq.(\ref{12}) is 
applicable in the low intensity limit for which the saturation parameter $S$ satisfies the inequality
 $S=2\Omega_{R}^{2}/(\Delta^{2}+\Gamma^{2})< 1$,  where $\Omega_{R}(0)$ is the Rabi frequency, defined below, evaluated at $z=0$,
and $\Gamma$ is the spontaneous emission rate.  At the high intensity regime,  corresponding to $S>>1$ one needs to adopt the 
 dressed atom approach [21].  The low intensity regime is the correct regime in the context of the atomic mirrors considered here
 for which it can readily be verified that $S<1$,  so that the dynamics based on the average radiation forces given by 
 Eq.(\ref{12}) is applicable.

Of the two average radiation forces, it is easy to see that
 ${\bf F}_{d}$ can act as a repulsive
force provided that
the detuning $\Delta_{0}$ is positive (blue detuning). Note that
${\bf v}$, the velocity of the atom, is in the plane of incidence as in Fig. 1, i.e. has only two
components, a z-component and a component parallel to ${\bf k}_{\parallel}$.  The Rabi frequency  $\Omega_{R}$ entering Eq.(\ref{12}) 
is defined for the electric dipole ${\bn {\mu}}$ in the evanescent field  ${\bf E}_{1}({\bf k}_{\parallel},{\bf r},t)$
\beq
\Omega_{R}=\left|\frac{\alpha{\bn {\mu}}{\bf .E}_{1}({\bf k}_{\parallel},{\bf r},t)}{\hbar}\right|
\eeq
where $\alpha$ is a complex amplitude factor such that $a\rightarrow \alpha$ in the classical electromagnetic 
field limit. It is in fact related to the 
intensity $I$ of the incident beam by the well known relation [23]
\beq
|\alpha|^{2}=\frac{IV}{\hbar c\omega}
\eeq
It can be seen from Eq.(4)
that the evanescent field possesses two vector components.  However, once the evanescent field has been set up, the average atomic
dipole moment vector at any given point aligns itself along, and follows the oscillations of, the local evanescent electric field vector.  The appropriate
Rabi frequency in this field-dipole orientation picture is thus given by [22]
\beq
\hbar
\Omega_{R}({\bf r})=\left|\alpha\right|\mu E_{1}({\bf r})
\eeq
where $\mu$ and $E_{1}$ are the magnitudes of these vectors.  Using Eqs.(4) and (9) we can write the square of the
Rabi frequency in the limit $\gamma\rightarrow 0$ in the following form,  emphasising the dependence on the angle of incidence $\phi$, which is related to $k_{\parallel}$ 
through
$c^{2}k^{2}_{\parallel}=\omega^{2}\epsilon_{2}\sin^{}\phi$,
\beq
\Omega_{R}^{2}(\omega,\phi,z)=\frac{4|\alpha|^{2}\mu^{2}k_{z2}^{2}c^{2}e^{-2k_{z1}z}}
{2\hbar V\epsilon_{0}\epsilon_{2}^{2}\omega}\left[
1+\frac{k_{z2}^{2}\epsilon_{1}^{2}}
{k_{z1}^{2}\epsilon_{2}^{2}}\left(\frac{\Lambda^{2}k_{z1}d}{\omega^{2}}-1\right)^{2}\right]^{-1}(1+k_{\parallel}^{2}/k_{z1}^{2})
\eeq
where $\Lambda$ is a convenient scaling frequency defined by
\beq
\Lambda^{2}=\frac{n_{s}e^{2}}{m^{*}\epsilon_{0}\epsilon_{1}d}
\eeq
with $d$ a convenient scaling length. Note the presence of the second term in the expression between the last pair of brackets in Eq.(17) 
which is proportional to $k_{\parallel}^{2}/k_{z1}^{2}$.  This represents the contribution from the z-component of ${\bf E_{1}}$,
 and since $k_{z1}$ can
be very small (when the angle of incidence is close to the total internal
reflection angle), this could lead to an enhancement,  in contrast to the contribution from the in-plane component 
(corresponding to the first term in the last pair of bracket in Eq.(17)) which does not exhibit this feature.  

Figure 2 displays the effects of varying the sheet electron density $n_{s}$ on the value of the squared Rabi frequency
 $\Omega_{R}^{2}(\omega,\phi,0)$, i.e. evaluated at the surface $z=0$.  This quantity provides an indication of the effectiveness
of the structure as an atomic mirror.  The curves correspond to different values of angle of incidence $\phi$.  There
are two features worthy of note here.  First,  for a given $\phi$ the variation of $\Omega_{R}^{2}$ with increasing $n_{s}$ 
is practically flat at the value corresponding to $n_{s}=0$ (absence of the metallic sheet) up to a point where it changes 
rapidly, increasing to a maximum and then dropping sharply to a relatively small value at a higher range of $n_{s}$.
This feature at large $n_{s}$ signifies screening effects.  The peak of $\Omega_{R}^{2}$ corresponds to the value of $n_{s}$
satisfying the condition $k_{z1}d=\omega^{2}/\Lambda^{2}$,  which is similar to (but not the same as) the dispersion
relation of the surface mode in this system which is [24-26]
\beq
\frac{k_{z1}d}{\left(\frac{\epsilon_{2}}{\epsilon_{1}}+\left[1-\frac{\omega^{2}}{c^{2}k_{z1}^{2}d^{2}}(\epsilon_{2}-\epsilon_{1})\right]^{1/2}\right)}=\frac{\omega^{2}}{\Lambda^{2}}
\eeq
This surface mode,  in turn, differs from the usual surface plasmon mode arising on a semi-infinite metallic half space.
 At an angle of incidence approximately equal to  $\phi_{0}$ for total internal
reflection,  the Rabi frequency exhibits a resonance which is at least two orders of magnitude larger than the value
corresponding to the absence of the metallic sheet.  At $\phi=\phi_{0}$ (not shown in the Fig. 2) the squared Rabi frequency $\Omega_{R}^{2}$ would be in the
form of a delta function. In principle, then, pronounced repulsion effects would be expected in conditions corresponding
to the region of the resonance.

The real mirror action is only partially influenced by the behaviour of the Rabi frequency; dynamical effects are,
of course, controlled by the forces acting on the atomic centre of mass. 
The atom experiences besides the average radiation force given by Eq.({\ref{12}), an attractive force due to interaction with the vacuum fields
 which are constrained by the presence of the surface.  At distances from the surface large compared to a reduced transition wavelength 
 $\lambda_{0}/2\pi$ the atom-surface force takes the Casimir-Polder form.  At distances 
smaller than $\lambda_{0}/2\pi$ the force assumes the van der Waals form.  As we show shortly, the important region for the atomic 
mirrors considered here, 
is, in fact, the van der Waals regime for which the force can be written as
\beq
{\bf F}_{{\rm vw}}(z)=-\frac{\partial U_{{\rm vw}}}{\partial z}
\eeq
where $U_{{\rm vw}}$ is the image potential.  As is well known, the van der
Waals potential arises as the position-dependent change in the radiative self energy 
due to the presence of the surface,  but it depends on the orientation of the 
electric dipole moment relative to the surface.  It should be emphasised that the evanescent field has no other role to play in the 
determination of the van der Waals potential,  except that it is responsible for 
the dipole orientation. The average electric dipole aligns itself parallel
to ${\bf {\hat e}}_{1}$, a unit vector in the direction of the local evanescent electric field so that 
${\bf E}_{1}({\bf r})={\bf {\hat e}}_{1}E_{1}({\bf r})$.  In the field-dipole
orientation picture this state of affairs applies at every point along the atom trajectory. The leading contribution to 
the van der Waals potential arises from the interaction of the electric dipole with its image and we therefore have 
\beq
U_{{\rm vw}}(z)=-\frac{\mu^{2}}{32\pi\epsilon_{0}z^{3}}\left\{3({\bf {\hat e}}_{1}
{\bf .{\hat z}})({\tilde {\bf {\hat e}}}_{1}{\bf .{\hat z}})-{\bf {\hat e}}_{1}{\tilde {\bf {.\hat e}}}_{1}\right\}
\eeq
where the unit vector   
${\tilde {\bf {\hat e}}}_{1}$ is  the image of ${\bf {\hat e}}_{1}$.  

The atomic reflection process is controlled by two separate
mechanisms.  Firstly, the spontaneous force ${\bf F}_{s}$ acts on the atom in the direction of ${\bf k}_{\parallel}$ and, secondly,
the combined force ${\bf F}_{d}+{\bf F}_{{\rm vw}}$ acts to repel the atom from the surface, decelerating its motion along the z-axis towards the surface,  
attaining an instantaneous halt before changing direction away from the surface at an appropriate turning point in the trajectory.
This behaviour can be seen more clearly by examining the corresponding repulsive potential 
$U_{T}({\bf v},z)=U_{{\rm vw}}(z)+U_{d}({\bf v},z)$ which is shown in Fig. 3 for a typical set of parameters.
Note that by controlling the average dipole direction  the field-dipole orientation picture influences the variations of both $U_{{\rm vw}}$ and $U_{d}$.

The trajectory of the atom of mass $M$ approaching a mirror for a given set up is obtainable by solving the equation of motion
\beq
M\frac{d^{2}{\bf r}}{d t^{2}}={\bf F}_{s}+{\bf F}_{d}+{\bf F}_{{\rm vw}}-Mg{\bf {\hat z}}
\eeq
subject to given initial conditions.  Figure 4 displays typical trajectories in the plane of incidence.  
The parameters are such that the spontaneous rate $\Gamma$ is taken to be the free space value $\Gamma_{0}$.  This is in fact a 
very good approximation in the trajectory region which is sufficiently far from the metallic sheet and the substrate.  The static detuning is 
taken to be $\Delta_{0}=5\times 10^{2}\Gamma_{0}$.  Finally the intensity of the light is assumed to be $I=2.0\times10^{4}$W m$^{-2}$.
It is straightforward to verify that for this light intensity and for the detuning value used here,  the saturation parameter
$S=2\Omega_{R}^{2}(0)/(\Delta^{2}+\Gamma_{0}^{2}) \approx 0.23$ which conforms with the low intensity regime.

In view of Fig. 4 one concludes that
the structure operates as an atomic mirror in three of the cases displayed,  while for the fourth case the trajectory 
of the atom terminates with a collision at the surface.  An approximate guide to the condition leading to 
a collision with the surface is to compare the maximum height $U_{max}$ of the potential in Fig. 3 to the initial kinetic energy
 $Mv^{2}_{z}(0)/2$.  For $v_{z}(0)>\sqrt{2U_{max}/M}$ a collision occurs.  This interpretation indeed conforms with the results
 of the type shown in Fig. 3.  Note that reflected atom trajectories are,  in general,  asymmetric with respect to the turning point.  This is a consequence 
of the action of ${\bf F}_{s}$ which in the present example where the light and the atom are incident on the same side (left side) 
of the z-axis, accelerates the atom to the right along the surface.  

In conclusion,  we have explored the influence of adding a metallic sheet to the 
usual evanescent mode atomic mirror set up.  The theory presented provides information about the range of metallic sheet densities and the  angle of incidence at which the
Rabi frequency exhibits pronounced enhancement effects.  We have also adopted the field-dipole orientation picture 
in which the average atomic dipole oscillates along the direction of the electric field of the evanescent mode at every point in the trajectory.
We have seen that this step which allowed us to identify the direction of the average dipole moment vector as that of the local electric
 field vector at every point in the trajectory, has important consequences for the evaluation of the forces and,  hence,  the dynamics of the atom.
Our results show that enhancement arises at an angle of incidence close to that of the total internal reflection
condition,  but requires a relatively high sheet density.  

\subsection*{Acknowledgement:}  JBK and CRB are grateful to the EPSRC for financial support.  This work
has been carried out under the EPSRC grant number GR/M16313
\newpage
 \section*{References}
\begin{enumerate}
\item V I Cooke and R K Hill, Opt.  Commun. {\bf 43}, 258 (1982)
\item W Seifert, C S Adams, V I Balykin, C Heine, Yu Ovchinikov and J Mlynek,  Phys. Rev. {\bf A49}, 3814 (1994)
\item G I Opat, S J Wark and A Cimmino, Appl.  Phys. {\bf B54}, 396 (1992)
\item S Tan and D F Walls,  J. Phys. II (France) {\bf 4}, 1879 (1994)
\item P Ryytty,  M Kaivola and C G Aminoff,  Europhys. Lett. {\bf 36}, 343 (1996)
\item  R J Wilson, B Holst and W Allison,  Rev. Sci. Instrum {\bf 70}, 2960 (1999)
\item D C Lau,  A I Sidrov, G I Opat, R J McLean, W J Rowlands and P Hannaford,  Eur.  Phys. J D {\bf 5}, 193 (1999)
\item R Cote, B Segev and M G Raisen,  Phys.  Rev. {\bf A58}, 3999 (1998)
\item L Cognet, V Savalli, G Zs K Horvath, D Holleville, R Marani, N Westbrook, C I Westbrook and 
A Aspect, Phys. Rev. Lett. {\bf 81}, 5044 (1998)
\item H Gauck, M Hartl, D Schneble, H Schnitzler, T Pfau and J Mlynek, Phys. Rev. Lett. {\bf 81}, 5298 (1998)
\item L Santos and L Rose,  Phys.  Rev.  {\bf A58}, 2407 (1998)
\item N Friedman, R Ozeri and N Davidson,  J. Opt. Soc. Am. {\bf 15}, 1749 (1998)
\item P Szriftgiser, D Guery-Odelin, P Desbiolles, J Dalibard, M Arndt and A Steane, Acta Phys. Pol. {\bf 93}, 197 (1998)
\item J P Dowling and J Gea- Banacloche,  Adv. Atom. Mol. Opt. Phy. {\bf 37}, 1 (1997)
\item A Landragin, J Y Courtois,  G Labeyrie, N Vansteenkiste, C I Westbrook and A Aspect,  Phys.  Rev.  Lett. {\bf 77}, 1464 (1996)
\item N Vansteenkiste,  A Landragin,  G. Labeyrie, R Kaiser, C I Westbrook and A Aspect,  Ann. Phys.-Paris {\bf 20}, 595 (1995)
\item S M Tan and D F Walls,  Phys.  Rev. {\bf A50}, 1561 (1994)
\item S Feron,  J Reinhardt,  S Lebouteux, O Gocreix, J Baudon, M Ducloy, J Robert, C Miniatura, S N Chormaic,
H Haberland and V Lorent, Opt.  Commun. {\bf 102}, 83 (1993)
\item T Esslinger,  M Weidenm\"{u}ller, A Hammerich and T W H\"{a}nch,  Opt.  Lett. {\bf 18}, 450 (1993)
\item J P Gordon and A Ashkin, Phys. Rev. {\bf A21}, 1606 (1980)
\item J Dalibard and C Cohen-Tannoudji, J. Phys. {\bf B18}, 1661 (1985) 
\item M Babiker and S Al-Awfi, Opt.  Commun. {\bf 168}, 145 (1999);
 L Allen,  M Babiker, W K Lai and V E Lembessis,  Phys. Rev. {\bf A54}, 4259 (1996)
\item R Loudon {\it The Quantum Theory of Light}, 2nd Edn. (Oxford, 1995)
\item F Stern, Phys. Rev.  Lett. {\bf 18}, 546 (1967)
\item  A L Fetter,  Ann. Phys. (NY) {\bf 81}, 367 (1973)
\item C R Bennett, J B Kirk,  and M Babiker, to be submitted.
\end{enumerate}
\newpage
\section*{Figure Captions}

\subsection*{\bf Figure 1}

Schematic arrangement of the elements comprising an evanescent mode atomic mirror with a metallic sheet.
The plane of incidence contains the internally reflected light beam as well as a typical atomic trajectory.  Here both 
the atom and the light are initially propagating on the same side of the vertical (z-axis).

\subsection*{\bf Figure 2}

Variation with areal density $n_{s}$ (in units of $n_{s}^{silver}=5.573\times 10^{18}$m$^{-2}$) of $\Omega_{R}^{2}$, the
squared evanescent mode Rabi frequency (in arbitrary units), evaluated just outside the metallic sheet in the limit of small $\gamma$.
The different curves correspond to different angles of incidence
$\phi = 42.00^{o}$ (dashes); $\phi = 47.00^{o}$ (dash-dots); $\phi = 52.00^{o}$ (dash-double dots).
The inset (solid curve) shows the variation for $\phi=41.37^{o}$, which is very close to the angle $\phi_{0}$ for total internal reflection.  
The parameters are such that 
$\epsilon_{2}=2.298$, corresponding to an angle of total internal reflection
$\phi_{0}=41.27^{o}$; $\Gamma_{0}=6.128\times 10^{6}$ s$^{-1}$; $\Delta_{0}=5.0\times 10^{2}\Gamma_{0}$ and
the intensity of the light is taken as $I=2.0\times 10^{4}$Wm$^{-2}$ for the Rb resonance at the transition wavelength $\lambda_{0}=780$nm.  
The scaling length $d$ is taken as $d=15$ nm.

\subsection*{\bf Figure 3}

Variation with $z$ (in units of $d=15$nm) of the combined static dipole potential $U=U_{d}+U_{{\rm vw}}$ 
acting on a Rb atom (in units of $\hbar\Gamma_{0}/2$).  The curves correspond to a fixed value of $n_{s}=215 n_{s}^{silver}$, 
but different values of angle of incidence $\phi$.
They are as follows: $\phi=42.00^{o}$ (dashes);  $\phi=45.00^{o}$ (dash-dots);  $\phi=47.00^{o}$ (dash-double dots);  
$\phi=41.37^{o}$ (solid curve).  The dotted curve shows the variation of the van der Waals potential.
In the evaluation of the potentials, the direction of the dipole moment vector 
conforms with the field-dipole orientation picture. The parameters are the same as those used in the evaluation of Fig. 2.

\subsection*{\bf Figure 4}

Atomic trajectories of a Rb atom in the atomic mirror arrangement shown in Fig.1 with the x-z plane as the plane of incidence.  
The angle of
incidence is fixed at $\phi=42.00^{0}$ and the metallic sheet corresponds to $n_{s}=215 n_{s}^{silver}$, as in Fig. 3.
In all cases the initial position of the atom is at the point $(x=0, z=100d)$,  where $d=15$nm. The different trajectories correspond to 
the same initial condition for the horizontal component of the velocity $v_{\parallel}(0)=0.054$
ms$^{-1}$,  but differ in their initial z-component of velocity $v_{z}(0)$.  They are as follows: $v_{z}(0)=0.1$ms$^{-1}$ (solid curve);
$v_{z}(0)=0.4$ms$^{-1}$ (dashed curve); $v_{z}(0)=0.7$ms$^{-1}$ (dash-dotted curve); $v_{z}(0)=1.1$ms$^{-1}$ (dash-double dotted curve).
The parameters are the same as those in Figs. 2 and 3.

}
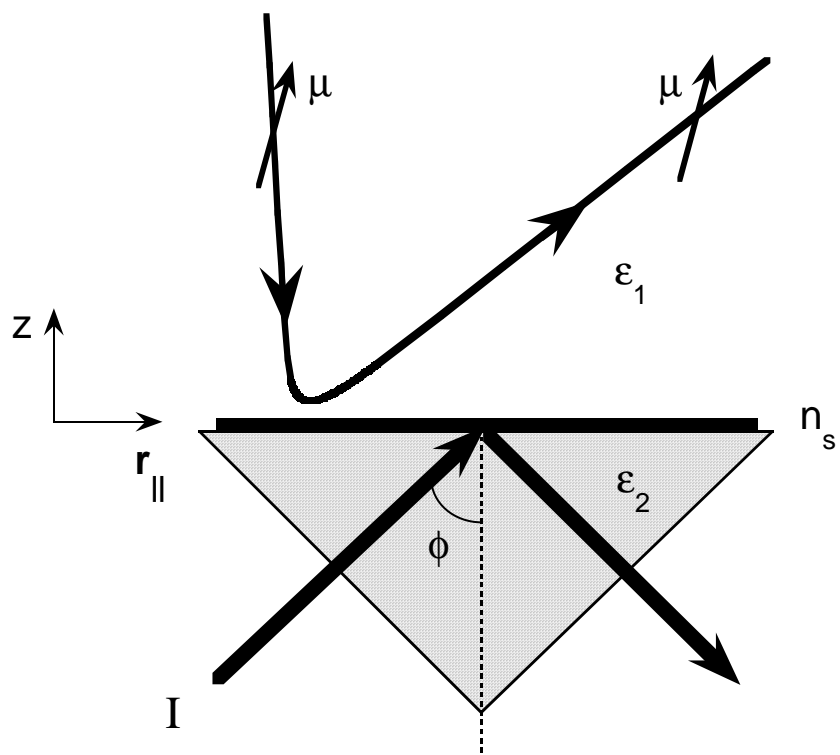
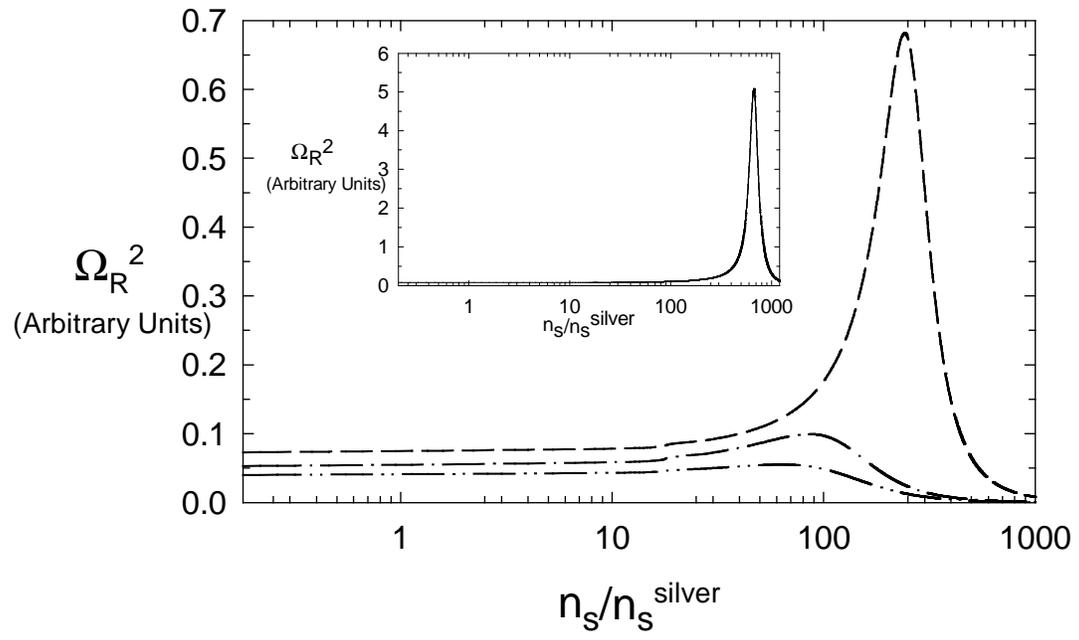
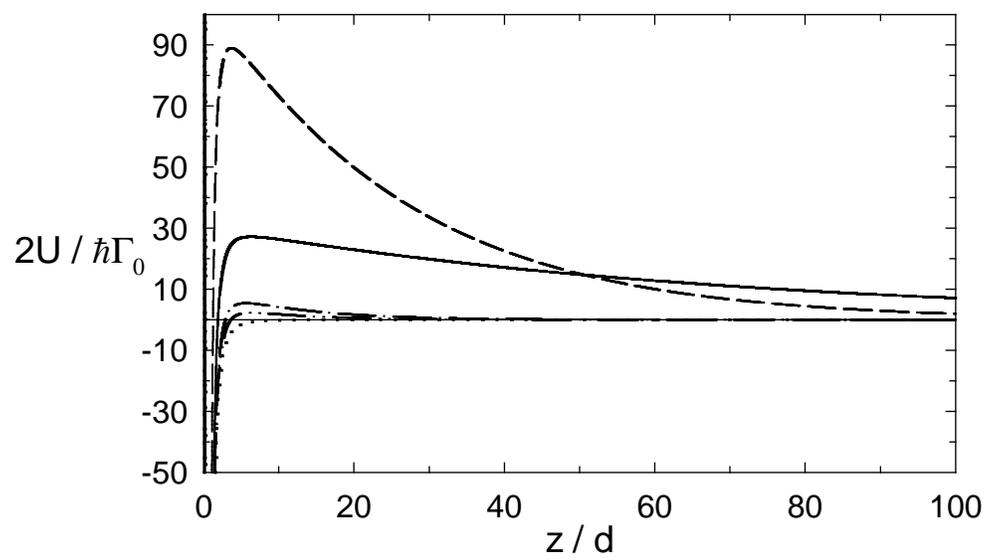
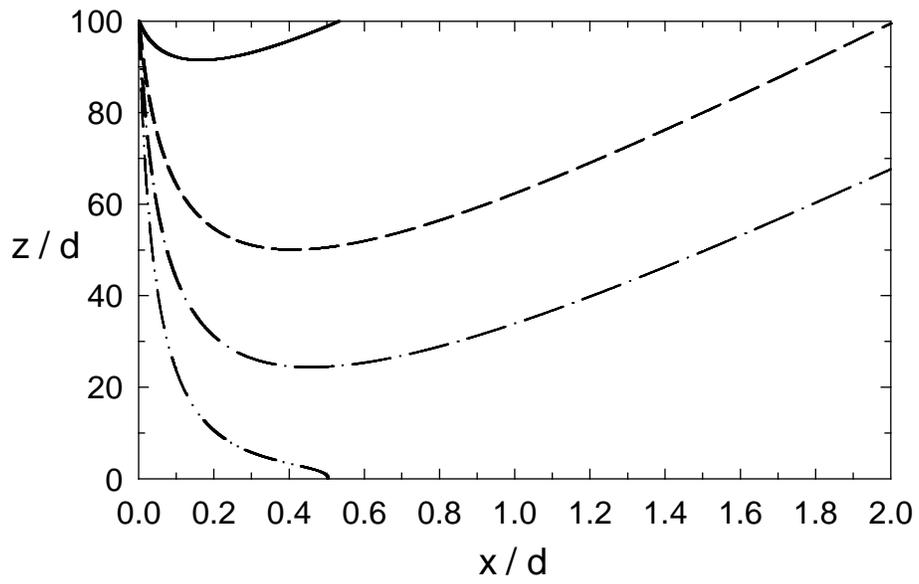
\end{document}